\documentclass[12pt,epsf,letterpaper]{article}
\usepackage{setspace}

\def\xI{x^I}
\def\xal{x^\alpha}
\def\bub{\mathcal{B}_0 }
\def\bubeps{\mathcal{B}_\epsilon }

\def\vecei{\vec{E} ^{(I)} }

\def\vecni{\vec{n} ^{(I)} }
\def\nai{n^{a(I)} }
\def\hatei{\hat{e}^{(I)} }
\def\ki{K^{(I)} }

\begin{document}

\begin{titlepage}

\begin{flushright}
\today
\end{flushright}

\vskip .5cm
\begin{center}
\baselineskip=16pt
{\Large\bf  Codimension Two Branes and}
\vskip 0.5cm
{\Large\bf   Distributional Curvature}
\vskip .5cm
{\large {\sl }}
\vskip 8.mm
{\bf Jennie Traschen} \\
\vskip 1cm
{

       Department of Physics\\
       University of Massachusetts\\
       Amherst, MA 01003\\
}
\vspace{6pt}
\end{center}
\vskip 1cm
\par
\begin{center}
{\bf Abstract}
 \end{center}
\begin{quote}
In general relativity, there is a well-developed formalism for working with the approximation
that a gravitational source is concentrated on a shell, or codimension one surface.
By contrast, there are obstacles to concentrating sources on surfaces that have a higher codimension,
for example,  a string in a spacetime with dimension greater than or equal to four. Here
it is shown that, by giving up some of the generality of the codimension one case,
curvature can be concentrated on  submanifolds that have codimension two. 
A class of metrics is identified such that  (1) the  scalar curvature and Ricci densities exist
as distributions with support on a codimension two submanifold, and (2)  using the
Einstein equation, the distributional curvature corresponds to a concentrated stress-energy with equation of state $p =-\rho$, where $p$ is the isotropic pressure tangent
to the submanifold, and $\rho$ is the
energy density.  This is the appropriate stress-energy to 
describe a self-gravitating brane that is governed by an area action, or a brane world 
deSitter cosmology. The possibility of having a different equation of state arise from 
a wider class of metrics is discussed.
 
\vfill
\vskip 2.mm
\end{quote}
\end{titlepage}

\section{Introduction}
 Working with sources that are concentrated
 on lower dimensional surfaces is an approximate description that is used
to advantage in analyzing many problems in classical physics, whether it is
modeling a section of plastic wrap as a two-dimensional membrane, an electric current as a line,
or a Newtonian mass as a point. In these examples, one locates the source by
a delta-function, which is well defined as a distribution in a fixed geometry. 
The situation is more complicated in general relativity, in which the geometry is
a dynamical field. And yet, there are problems of interest in which
sources for the gravitational field  are naturally viewed as being concentrated on a submanifold--
cosmic strings, branes in string-motivated gravity, and brane-world cosmologies.
 codimension one branes in general relativity work fine. The Israel junction
  formalism \cite{Israel:1966rt}
gives a prescription for constructing a solution to the Einstein equation with a shell source.
Instead of being smooth, the metric is only continuous across the shell, and has a jump
in its normal derivative, which is interpeted via the Einstein equation as due to
a shell of stress energy.

However if a brane is self-gravitating, and is concentrated on a lower dimensional surface,
then problems arise. 
There is no prescription for sources concentrated on 
submanifolds with codimension greater than one. Indeed, in reference \cite{Geroch:1987qn}
it was shown that for metrics that are well enough behaved so that the Riemann tensor
exists as a distribution, in general curvature can only be concentrated on codimension one
surfaces. In the work presented here, by ``giving up a little",
 and assuming some added structure on the spacetime,
we present a construction that is tailored to allowing concentrated curvature
on a codimension two  surface.  

This work was motivated in part by known analytic solutions 
that {\it do} describe stress energy concentrated on a codimension two surface. The
most famous example is $3+1$ dimensional flat space minus a wedge, which is
a model for the spacetime outside a straight cosmic string. In this picture, the stress-energy
of the string is concentrated on a codimension two submanifold, the $1+1$ dimensional axis of
the string. This simple model matches the properties displayed by the solution
for a finite width cosmic string composed of gauge and scalar fields \cite{Garfinkle:1985hr}. Outside the
core of the string, the metric approaches flat space minus a wedge exponentially fast. 
Inside the core, to leading order the equation of state of the matter is $p=-\rho$, 
where $p$ is the pressure tangent to the string, and $\rho $ is the energy density. An example
of a finite width brane cosmology is given in reference \cite{Rubakov:1983bb}.

 A second analytic example is provided by static Kaluza-Klein  Killing bubbles
 \cite{Elvang:2002br} \cite{Kastor:2008wd}. A Killing bubble is a 
 minimal surface that arises as the fixed surface of a spacelike Killing field. 
 In these solutions the bubble, at fixed time, is a $(D-3)$-dimensional
 sphere, where $D$ is the spacetime dimension.  The  static Kaluza-Klein bubble metric has the form
 \begin{equation}\label{kkmetric}
 ds^2= -dt^2 + f(R) k^2 d\phi^2 +{1\over f(R)} dR^2 +R^2 d\Omega ^2 _{D-3} 
 \end{equation}
 where $ f(R) =1-R_0 /R $, $0\leq \phi \leq 2\pi$ and $0<k\leq 1$. The Killing vector
  $(\partial / \partial \phi )$ vanishes at $R=R_0$, which is a minimal $(D-3)$-sphere 
  with non-zero area. This is the Killing bubble. 
   The two dimensional space orthogonal to the bubble, here the
   $R-\phi$ plane,  is generally taken to be smooth, which requires $k=1$. To see this,
   one expands the metric near the bubble, 
  \begin{equation}\label{kkmetrictwo}
 ds^2\rightarrow -dt^2 + k^2 y ^2  d\phi^2 +dy ^2 +R_0 ^2 d\Omega ^2 _{D-3} 
 \end{equation}
For $k=1$, the spatial geometry has the form of a smooth ${\cal R}^2$ orthogonal to the
  bubble. With the Kaluza-Klein  boundary conditions on the metric (\ref{kkmetric}), this
  is usually described as a cigar cross a sphere, and the minimal sphere
 is at the tip of the cigar. However, the solutions make sense even if the ${\cal R}^2$
 has a missing angle with $k<1$. Then the smooth cigar is replaced by
 a cone with the bubble at the tip. The missing angle geometry
 may be interpeted as a $p=-\rho$ source that wraps the minimal sphere \cite{Kastor:2008wd}, analogous to the idealized cosmic string or a Euclidean black hole vortex \cite{Dowker:1991qe}.
  Metrics having the same limiting form as (\ref{kkmetrictwo}) have been exploited in building  brane world cosmologies with two extra dimensions \cite{Carroll:2003db}.
  
  Therefore, on  one hand,  the picture that a metric locally of the form
   (\ref{kkmetrictwo}) describes matter concentrated on the $D-2$ submanifold
  at the tip of the cone $y =0$, has been used in many models. On the other hand,
  the result of  \cite{Geroch:1987qn} tells us that in general, the Riemann tensor of  such a
  metric is not well defined as a distribution.
  The construction here aims to resolve these statements, and to
  generalize the situation with the known solutions. We identify curvature invariants of the
  spacetime that $are$ well defined as distributions in the codimension two case. This is 
  sufficient to work out the effective distributional stress energy, without having to
 use a smooth fill-in, such as was done for a straight cosmic string using the Abelian Higgs model
 \cite{Garfinkle:1985hr}. 
 
  We will proceed as follows.
Let $\bub$ be a codimension two submanifold
on which curvature is to be concentrated, and assume that locally there is a codimension
two foliation of surfaces $\bubeps$ that includes $ \bub$. Each
 surface has two normal forms,   which are used to
split the spacetime into surfaces ``tangent to" $\bub$, and ``normal to" it. The singular
behavior is confined to the metric on the two-dimensional  surfaces normal
to $\bub$. We specify
conditions such that (1) the spacetime scalar curvature density, and the Ricci density, exist
as distributions concentrated on the codimension two submanifold, and (2)  using the
Einstein equation, the concentrated curvature corresponds to a concentrated stress-energy with equation of state $p=-\rho $. The conditions needed
are summarized in the preface to equation (\ref{intcurv}). 

The assumptions made are simple ones, and yield a family of metrics that includes
the analytic examples, but is more general as there are no symmetry  requirements on the metric.
Necessary and sufficient conditions are given in Appendix 1 for
the metric to have the assumed form, which is a particular block diagonal structure.
 Further, the calculation elucidates how this form of the metric implies that
 the equation of state on the brane is $p=-\rho$. This is an issue of interest in the context of 
 brane world cosmologies -- one would like the brane to contain other types of stress-energy
 as well. Various approaches to this have been studied, including adding  higher derivative
 terms   \cite{Cline:2003ak}, and adding a thickness to the brane \cite{Peloso:2006cq}. In our concluding section, we discuss
 generalizations that may yield other equations of state on the brane.
 
The construction presented here has similarities to stress-focusing that occurs in 
crumpled membranes \cite{TAWitten}.
If external forces are applied to a thin material, the membrane bends in response.
 However, if one crumples the membrane, forcing it to occupy a smaller region, then one or
 more vertices appear at which stretching of the material occurs. Bending generates 
extrinsic curvature, which is cheaper energetically, whereas stretching corresponds to
focused intrinsic curvature. This is more expensive energetically, but is the only option
under certain external forces. In the gravitational construction here, the extrinsic curvatures 
are assumed to be well behaved, but focused curvature occurs in the two-dimensional
space normal to the brane. An explicit example of crumpling in the gravitational context, 
with codimension two vertices connected by codimension one ridges, has been 
constructed in \cite{Kaloper:2004cy}.

The plan of this paper is as follows.
Section 2 sets up  features of the  geometry near $\bub$. These properties are used in Section 3 to
compute the scalar curvature  and Ricci tensor densities, to show that they 
are concentrated on $\bub$, and to derive the equation of state for the concentrated
stress-energy. Section 4 briefly discusses the next step of finding solutions to 
the vacuum Einstein equation with concentrated sources, and the role played
by minimal surfaces.
Section 5 mentions some open questions. Some technical points are deferred
to the appendices.
We use the convention that latin letters run over all values $a, b=0, ...,D-1$. Greek indices
run over $D-2$ coordinates tangent to the surface, $\alpha, \beta, = 0,3,..., D-1$, while
$i,j= 3,..., D-1$ denote spatial coordinates tangent to the surface.
Capital roman letters index the two normal directions $I,J=1,2$

\section{Geometry Near the Bubble}
Let a spacetime $\cal{M}$  contain  a codimension two spacelike submanifold $\bub$ on which
curvature is to be concentrated. We will refer to $\bub$ as ``the bubble", since the Killing
bubbles referred to in the introduction provide analytic examples of the more
general construction developed here.
Assume that the bubble arises as the
intersection of the level surfaces of two smooth functions, $\xI =0 , I=1,2$. Let  $ x^\alpha ,\alpha =
0,3,..., D-1$ be a choice of $D-2$ other  good coordinates on  $\cal{M}$
in a neighborhood of $\bub$. Since the analysis in this section applies to a neighborhood
of $\bub$, we will not keep repeating this phrase, but that will be understood. The
forms $\underline{n}^{(I)} =\underline{dx}^I $ at $\xI =0$ are two normals to $\bub$, though
not necessarily unit. We also assume that the family of codimension two surfaces $\bubeps$,
defined by $\xI =\epsilon ^I$, is a foliation  by smooth spacelike submanifolds.
Let $B_{ab} (\xal , \xI )$, evaluated at $\xI =\epsilon ^I$, be a smooth family of metrics on the $\bubeps$. The coordinate vectors
 $\{ {\partial\over\partial x^\alpha} \} $ are a basis for the tangent space 
   $\cal{T}(\bubeps )$, and the forms $\underline{dx}^{(\alpha)}$ are a basis for the dual space
    $\cal{T}^* \bubeps$ of the submanifolds. 
     
  We split the spacetime metric as
\begin{equation}\label{metricnear}
g_{ab} =B_{ab} +\sigma_{ab}
\end{equation}
with $\sigma _{ab} B^{bc} =0$. Since the curvature
 is nonlinear in the metric and its inverse, one has to start with a metric field that is better behaved
 to end up with distributional curvature. So the metric $g_{ab}$ itself is not
 a distribution, that is, it is not already concentrated on a submanifold. 
 Specifically,  $B_{ab}$ is assumed to be smooth,
 and $\sigma_{ab}$ is a tensor field that is smooth in any region bounded away from
 $\bub$. The value of $\sigma_{ab}$ at $\bub$ is given by its limiting value as $x^I \rightarrow 0$,
 which may be zero or infinity. This is different from the assumption  made in reference \cite{Geroch:1987qn} that the metric is regular. Confining the singular behavior to the
 normal plane is similar to the ``normal dominated singularity'' approach of Israel,
 in analyzing line sources in four dimensions \cite{Israel:line}.

 We require that the divergence is mild enough
  that local volumes  are finite:  Let $V_l$ be a $(D-1)$-dimensional
spatial volume that contains all or part of $\bub$ , so $V_l :\{ 0\leq x^I \leq l^I  , 0\leq x^j \leq x^j _{max} \}$. Let $g^{(D-1)}_{ab}$ denote the spacetime metric restricted to $V_l$.
  Then we shall assume that the volume of $V_l$ is finite,
\begin{equation}\label{finitevol}
\int _{V_l} \sqrt{g^{(D-1)} } <\infty
\end{equation}
  and that for any smooth function $F$
integrated over $V_l$, then in the limit $l^I\rightarrow 0$, this integral vanishes:
\begin{equation}\label{intvanish}
\lim_{l\rightarrow 0} \int _{V_l} \sqrt{g^{(D-1)} } F =0
\end{equation}
This latter condition rules out the volume element itself acting as a delta-function.

The simplest construction is when the spacetime can be foliated by a second family of 
two dimensional submanifolds $N$ which are normal to the $\bubeps$, with $\sigma _{ab}$
the metric on $N$. In this paper we will  assume this holds, $and$ that the spacetime metric
 is locally block diagonal. Hence
\begin{equation}\label{blockmetric}
 ds^2 = \sigma_{IJ}dx^I dx^J   +B_{\alpha \beta}dx ^\alpha dx^\beta
 \end{equation}
The two dimensional metric $\sigma_{IJ}$ can always be written locally in conformally flat
coordinates,
\begin{equation}\label{conformalmetric}
d\sigma ^2 =\sigma _{IJ} dx^I dx ^J = \Omega ^2 (\delta _{IJ} dy^I dy ^J ) = \Omega^2 (dr^2 
+r^2 d\phi ^2 )
\end{equation}
with $r^2 =\delta _{IJ} y^I y ^J$ and $0\leq \phi \leq 2\pi$. The simplification that
we have gained is that the conditions for distributional curvature on the bubble
can be stated in terms of conditions on the single function $\Omega$. 
We will see that curvature can be concentrated on $\bub$ for appropriate choice 
of $\Omega$. From the details of the arguments below, we expect that it is possible to
generalize to a non-block diagonal metric, but still retaining the split (\ref{metricnear}).

When does one expect the block diagonal form to hold?
 Let $g$ be a metric that is smooth in any neighborhood not including $\bub$ .  Choose  positive $\epsilon ^I$. In Appendix I we show that  a necessary and
sufficient condition  for the existence of coordinates  so that $g$ can be put in  block form (\ref{blockmetric}) in a neighborhood of $\bubeps$ is dictated by Frobenius' Theorem. 
The condition is that the commutator of the normal vector
 fields $\vec{n}^{(I)} =g^ {aI}{\partial\over \partial x^a }$ closes, see equation (\ref{commute}).
 Therefore we need that the $\vecni$ commute in a neighborhood 
 $0<\epsilon ^I \leq \epsilon ^I _0$, for some $\epsilon ^I _0$, for the metric to be able to
 be put in the form (\ref{blockmetric}).
 
Lastly, we note that with the assumption that
 $\sigma_{ab}$ is the metric on a submanifold $N$, then the finite volume condition (\ref{finitevol})
can be stated in terms of finite areas of $N$.
 Let $D_l$ be the disc $0\leq x^I \leq l^I$ located at some point  $x^\alpha $
on the minimal surface. We require that 
\begin{equation}\label{finitearea}
A(D_l )=\int _D \sqrt{\sigma} \   < \infty
\end{equation}
and that $A(D_l )\rightarrow 0$ as $l^I\rightarrow 0$. 
If $\Omega ^2\rightarrow r^{-2\mu} $, then finite area requires $\mu < 1$.

\section{Curvature }
The strategy for concentrating curvature on the codimension two minimal surface is
to isolate the singular behavior in the two dimensional metric $\sigma_{IJ}$, that is, in
the conformal factor $\Omega$. $\Omega$ can be chosen in such a way that
 appropriate spacetime curvature tensor densities are well defined as distributions with support on
 $\bub$.  To do this, we relax the assumptions of reference \cite{Geroch:1987qn} on the spacetime metric.
 There $g_{ab}$ was defined to be a $regular$ metric in a region  if (i) it and its inverse exist 
 everywhere, (ii) they are locally bounded, and (iii) the weak first derivative of $g_{ab}$ 
 exists and is square integrable. These conditions were chosen to
  ensure that the Riemann tensor is defined as a distribution;
  that is, for any smooth tensor $density$ $s^{abcd}$, the integral $\int R_{abcd} s^{abcd}$
  exists. Further, the outer product of the curvature tensor and the metric are
  distributions, and hence the usual contractions of Riemann are as well. 
  
  Reference \cite{Geroch:1987qn} also notes that
  under weaker conditions on the metric, particular curvature invariants can be defined
  as distributions. An example given is two-dimensional flat space minus a wedge,
  for which the conformal factor and scalar curvature densities of $\sigma_{ab}$ are 
  \begin{eqnarray}\label{cone}
  \Omega ^2  &=r^{-2\mu}\ , \   \mu < 1 \\  \nonumber
  \sqrt{\sigma} R[\sigma ] 
  &=4\pi \mu \delta ^{(2)} (y^I  )
\end{eqnarray}
 For $0< \mu < 1$ this is the metric of a cone,
while if $\mu <0$ the curvature is negative--there is extra angle. For brevity,
we will refer to this metric as a cone in either case.  $\Omega$ 
satisfies the finite area condition as long as $\mu <1$.  It turns out that the sign
of $\mu$ determines the sign of the energy density on $\bub$.
 
 Now, although $\sigma$ is not regular (unless $\mu =0$), its scalar curvature density 
   is the familiar, flat space, two-dimensional delta function, and
  hence makes sense as a distributional density.  
  However, not all the curvature invariants are well defined. $R[\sigma ]$ is zero as a distribution,
as it gives zero  integrated against a smooth tensor density, and the Riemann tensor is infinite.
So we will proceed to  make use of the well- behaved features of the two-dimensional curvature density,
but as part of a higher dimensional spacetime. The way that the singular two-dimensional
metric  contributes to the full spacetime curvature is controlled by splitting the spacetime metric into
submanifolds  parallel to $\bub$, and orthogonal to it. A different approach  was taken
 in \cite{Garfinkle:1999xv}. Here the Riemann tensor is defined as a distribution for a
 wider class of ``semi-regular" metrics,  which includes flat space minus a wedge.

We start with the Gauss-Codazzi relations for smooth metrics. 
Splitting the metric as in (\ref{metricnear}), the various projections of the Riemann tensor
for $g$ can be written in terms of the Riemann tensors for $B$ and $\sigma$, plus
extrinsic curvature terms. The needed details are given in (\ref{gcrels}).
The spacetime scalar curvature is
\begin{eqnarray}\label{curvature}
 R[g] &= & R[\sigma ] +R[B] +  \Sigma_I [ (K^{(I)} )^2 - K^{(I)} _{ab}  K^{(I)ab}] \\ \nonumber
&&  + \lambda_a \lambda^a
  -\lambda_{abc} \lambda^{abc} +2(\nabla_a \pi^a -\nabla_a \lambda^a )
 \end{eqnarray}
Here $\ki _{ab}, \pi ^a $, and $\lambda_{abc}$ are extrinsic curvature tensors; the general definitions are 
given in equations (\ref{bigextrcurv}) , (\ref{pitok}).
 For a metric of the block diagonal form (\ref{blockmetric}), (\ref{conformalmetric}), one finds 
\begin{equation}\label{kandlambda}
\ki _{\alpha\beta}= {1\over 2\Omega} \partial_I B_{\alpha\beta} \quad
\mathrm{and} \quad \lambda _{IJ\alpha}=\sigma_{IJ} {\partial _\alpha\Omega \over \Omega}
\end{equation}

The decomposition of the scalar curvature in (\ref{curvature}), and of the Ricci tensor in 
(\ref{project}), are derived assuming the spacetime metric is smooth. Here, we 
proceed to use these equalities as true when integrated. That is, if the integral of the
right hand side is finite, we equate that to  the integral of the left hand side, as in
equations (\ref{intcurv}) and (\ref{stressenergy}) below.

Substituting the extrinsic curvatures into the terms of $R[g]$ gives 
\begin{eqnarray}\label{curvaturetwo}
 \Sigma_I [ (K^{(I)} )^2 -   K^{(I)} _{ab}  K^{(I)ab}]& = &
  {1\over 4 \Omega ^2 } \Sigma_I  [ (B^{\alpha\beta}\partial _I B_{\alpha\beta} )^2
  -B^{\alpha\mu}  B^{\beta\nu} (\partial _I B_{\alpha\beta} )(\partial _I B_{\mu\nu} )] \\ \nonumber
   \lambda_a \lambda^a -  \lambda_{abc} \lambda^{abc} &=  &
    2B^{\alpha\beta} \Omega ^{-2} \partial _\alpha \Omega \partial _\beta \Omega  \\ \nonumber
    \nabla_a \pi^a -  \nabla_a \lambda^a  &= &
    {1\over \ \sqrt{B}   \Omega ^2 } \left[    
   - \delta^{IJ} \partial _I(\sqrt{B} B^{\alpha\beta})\partial _J B_{\alpha\beta})
   +2   \partial _\alpha \left(\sqrt{B}\Omega B^{\alpha\beta}\partial _\beta \Omega ) \right) \right]
    \end{eqnarray}
   Next, integrate $R[g]$ over $V_l$. In the limit that $l^I\rightarrow 0$, since volumes  are locally
finite, any bounded term on the right hand side of (\ref{curvaturetwo}) will give zero. 
Also, any term in the integrand that is the form $\Omega ^{-2} \times smooth$
will give zero upon integration, since
$\sqrt{\sigma} = r\Omega ^2$.
Since $B_{\alpha\beta}$ is assumed to be smooth, $R[B]$ does not contribute. 
 The terms that  come from $\lambda _{abc}$ depend on
 $\Omega ^{-2}B^{\alpha\beta}\partial _\alpha \Omega\partial _\beta\Omega $ and  
 $\Omega ^{-2}B^{\alpha\beta}\Omega\partial _\alpha \partial _\beta\Omega$,
  and so far we have not made
 any assumptions about this behavior. Different assumptions may give different 
 results of interest. The choice that we make here is that these terms are bounded. It will be shown
 below that this choice 
 will give a $p=-\rho $ equation of state for the concentrated curvature.
 
{\bf Summarizing:} Let the metric near $\bub$ have the block diagonal form
(\ref{blockmetric}), (\ref{conformalmetric}). $B_{ab}$ is assumed to be smooth. $\sigma _{ab}$
is smooth  in any region not including $\bub$. $\sigma _{ab} $ may approach zero or infinity on $\bub$,
but volumes are finite, as stated in (\ref{finitevol})  or (\ref{finitearea}). Assume that
 $\Omega^{-2} B^{\alpha\beta}\partial _\alpha \Omega\partial _\beta\Omega \  $ and  
 $\Omega^{-2} B^{\alpha\beta}\Omega\partial _\alpha \partial _\beta\Omega$
  are bounded (including on $\bub$). Lastly, assume that $|r\partial_I \Omega | < r ^{1+\epsilon}$
  with $\epsilon > 0$. This last assumption is needed for deriving (\ref{pandrho}) below.
  
  All of these assumptions are satisfied by the cone metric with a position dependent amplitude,  $\Omega =S(x^\alpha )r^{-\mu}$ with $\mu <1$ .
  
Let  $V_l$ be a $(D-1)$-dimensional
spatial volume that contains all or part of $\bub$ as above,
 $V_l :\{ 0\leq x^I \leq l^I  , 0\leq x^j \leq x^j _{max} \}$. Integrate
the scalar curvature (\ref{curvaturetwo}) over
$V_l$ and take the limit as $l^I \rightarrow 0$, which means that the spatial volume
collapses to the codimension two surface $\bub$.
 Let $A(\bub )$ be the area of $\bub$ if it is compact, otherwise $A$ is  the area
 of some subset. Then
\begin{eqnarray}\label{intcurv}
\lim_{l\rightarrow 0} \int _{V_l} \sqrt{g^{(D-1)}} R[g]  & = &A(\bub ) \lim_{l\rightarrow 0}\int _{D_l}
 \sqrt{\sigma} R[\sigma ] \\ \nonumber
& =&-2A(\bub ) \lim_{l\rightarrow 0} \int dy^1 dy^2 \nabla ^2 ln\Omega
 \end{eqnarray}
 where we have substituted the expression for the two-dimensional 
curvature $R[\sigma ]= -{2\over\Omega^2}\nabla ^2 ln\Omega$. 
For the cone metric this becomes
\begin{equation}\label{intcurvcone}
\int _{V_l}  R[g] \rightarrow 4\pi \mu A(\bub )
\end{equation}

Hence the scalar curvature of the spacetime metric has support on the codimension two surface
$\bub$. Equations (\ref{intcurv}) and (\ref{intcurvcone}) are
one of the main results of the paper.

An effective stress-energy associated with this concentrated curvature is defined
by integrating components of the Einstein tensor over $V_l$ and letting $l\rightarrow 0$.
Let $u^a$  and $x^a$ be  unit timelike and spacelike vectors that are tangent to $\bub$. 
 So $B^a _b u^b = u^a$ and $B^a _b x^b = x^a$.
(There are $D-3$ independent such spacelike vectors, but here we practice index suppression.)
Define an effective stress-energy concentrated on the surface by
\begin{eqnarray}\label{stressenergy}
8\pi \rho =\lim_{l\rightarrow 0} \int _{V_l} u^a u^b G_{ab} = 
\lim_{l\rightarrow 0} \int (u^a u^b R_{ab} +{1\over 2}R) \\  \nonumber
8\pi p_x =\lim_{l\rightarrow 0} \int _{V_l} x^a x^b G_{ab} = \lim_{l\rightarrow 0} \int (x^a x^b R_{ab} -{1\over 2}R)
 \end{eqnarray}
 and similarly for the other components.  The Ricci and scalar curvatures are
 of the spacetime metric $g_{ab}$.
  
So the next step is to compute the components of the Ricci tensor tangent to the surfaces $\bubeps$,
$B^m _a B^n _c R_{mn}=B^m _a B^n _b (B^{bd}+ \sigma ^{bd} )
 R_{mbnd}[g]$. Using the first two  equations of (\ref{gcrels})  these terms become
 \begin{eqnarray}\label{project}
 B^m _\alpha B^n _\gamma B^{b d} R_{mb nd}[g]  &
 = & R_{\alpha \gamma }[B] +\Sigma_I \left(  K^{\beta (I)} _\alpha  \ki _{\gamma \beta}
  -\ki \ki_{\alpha \gamma}  \right) \\  \nonumber
  B^m _\alpha B^n _\gamma  \sigma^{b d} R_{mb nd}[g] & = & \Sigma_I \left(
    K^{\beta (I)} _\alpha  \ki _{\gamma \beta} \right) \\ \nonumber
& & + B^\mu _\alpha B^\beta _\gamma \left( - 2\Omega ^{-1} \partial _\mu \partial _\beta \Omega  +
2\Gamma ^\lambda _{\mu\beta} \Omega ^{-1}  \partial _\lambda \Omega
  -\Omega ^{-2} \delta^{IJ} \partial _J \partial _I B_{\mu\beta}         \right)
 \end{eqnarray}
 where $\ki _{\alpha\beta}$ is given in (\ref{kandlambda}).
 
Integrate these equations over $V_l$ and take the limit $l\rightarrow 0$. Then
 under the assumptions stated for (\ref{intcurv}), all of the terms on the right hand side of (\ref{project})
 give zero. Hence in computing the stress-energy (\ref{stressenergy}) on $\bub$ 
 \begin{equation}\label{pandrho}
 8\pi \rho =-8\pi p_i =\lim_{l\rightarrow 0} \   {1\over 2} \int _{V_l} R[g]  
 =A(\bub ) \lim_{l\rightarrow 0} \   {1\over 2} \int _{D_l} R[\sigma ]  
 \end{equation}
 When the normal plane has a cone metric with  $\Omega =S(x^\alpha )r^{-\mu}$ this become
 \begin{equation}\label{pandrhocone}
 \rho=-p ={1\over 4}\mu A(\bub )
 \end{equation}
  
 Equations (\ref{pandrho}), (\ref{pandrhocone}) giving the equation of state of the concentrated curvature   is the second main result of this paper.

  What we have learned  is that  with the conditions stated for (\ref{intcurv}),
  the Ricci tensor  projected
 tangent to $B_{ab}$ does not contribute to the integrated Einstein tensor,
 as the volume collapses to the surface.  Only the scalar
  curvature term contributes, giving an effective equation of state $p=-\rho$.
 This is reminiscent of the way that an effective cosmological constant (which has  $p=-\rho$)
 arises from a scalar field when the Lagrangian is potential dominated. This result
 is very different from the codimension one case. Shells can have any effective
 stress-energy, by appropriately choosing the jump in the extrinsic curvature across
 the shell. So, the question arises whether it is possible to get other types of
 stress-energy on a codimension two brane \cite{Cline:2003ak} \cite{Peloso:2006cq}.
 To get a different equation of state, the Ricci tensor must contribute to the
 concentrated stress energy in (\ref{stressenergy}). We defer further discussion to
 the  Section 5 on open questions.
  
Actually, it remains to check the other projections of the Einstein tensor. It turns out that 
 one additional type of term arises ,
 $ \Omega ^{-1} (\partial _I \Omega / \Omega ) (\partial _\alpha \Omega / \Omega )$.
 This integrates to zero as  $l^I\rightarrow 0$, if $| r\partial _I \Omega| < r^{1+\epsilon}$. Hence with the
same assumptions as for (\ref{intcurv}), these other components of the Einstein equation
 integrate to zero in the limit $l^I\rightarrow 0$.
 
  A couple of details are worth mentioning,
 as  they may be relevant to possible extensions to higher codimension.
The most singular piece comes from the $\sigma _a ^m \sigma_a ^n G_{mn}[g]$ projection,
which includes $R_{ab} [\sigma ] -1/2 \sigma _{ab} R[\sigma ]$. However, for $\sigma _{ab}$
two-dimensional, this term vanishes, which is special for two dimensions. 
The next most singular piece
comes from the cross terms of the Ricci tensor $\sigma _a ^m B_a ^n R_{mn}$. 
Reference \cite{Gergely:2005sd} was used to analyze this term. These
mixed components contain the term discussed in the previous paragraph. 
There is one other potentially unbounded term,
$\sigma _a ^p \sigma _c ^m \nabla _p \pi _m =
 -\Sigma_I \sigma _a ^p \sigma _c ^m \ki \nabla _p \hatei _m $.
 Simple power counting implies that the term can be mildly divergent, although
  it integrates to zero with the stated conditions . However, using the fact that the
  commutator of the $\hatei$ closes, one can check that  this term is  actually finite, and  is zero 
  if the commutator vanishes\footnote{ Of course, this statement only makes sense if the
   unit basis forms $\hatei$ are defined, which may not be true on the bubble.
However, since the term only involves first
 derivatives of $\Omega$, it can not contain distributional curvature; that is, the value of its 
 integral over $V_l$  is independent of  the value on $\bub$.}.
 So it may be that geometrical criteria also tame some of the other terms.
 
 \section{Minimal Surfaces and Solutions }
 
 The focus of 
 this paper has been to identify a class of metrics that describe
curvature concentrated on codimension two surfaces. It turns out that a set of natural
assumptions imply that the concentrated source has an equation of state $p=-\rho$.
Though we have  not addressed the questions of finding solutions
to the  vacuum Einstein equation, one expects that on solutions $\bub$ 
will be a minimal surface. This is because a $p=-\rho$ test brane
 is governed by an area action, and propagates as a minimal surface.
  So it is reasonable to guess that
if this sort of brane is self-gravitating, a consistent configuration would be for the location
of the brane to be a minimal surface. For example, in
the two analytic solutions discussed in the introduction,   $3+1$ dimensional flat space minus a wedge and the spherical
 bubble with a missing angle, $\bub$ is a minimal surface.
 In both examples, the effective stress energy is of
 the form $\mu B_{ab}$ on $\bub$, and the spacetime is vacuum elsewhere.  To look for other solutions,
 one would need to pick a manifold and  $\bub$, then
 solve the vacuum Einstein equation, with the metric having the allowed singular behavior near $\bub$.
 Our assumption that the metric is block diagonal limits the class of solutions, but
 hopefully that assumption will be relaxed in future work.
 
This raises the question of whether a surface $\bub$ can be described as ``minimal", if
 the geometry is allowed to be singular as in (\ref{intcurv}).
 Hence we close  by showing that even though $\sigma _{ab}$ is allowed to be mildly singular, 
 the description of $\bub$ as a minimal surface does  make sense.
 
  A minimal  surface
has the property that if it is deformed, the area does not change to first order.
 Since the area is unchanged under tangential 
deformations, only variations off the original surface must be considered. The usual
procedure is to compute $\delta A$ under deformations in the directions $\nai$. In this paper
the normal forms have been defined so that they are
 well behaved, but the vectors $\nai =g^{ab}n^{(I)} _b$ may diverge or
be zero on $\bub$. However, the coordinate vector fields $({\partial \over \partial x^I})$
are smooth, and have an off the surface component, as $({\partial \over \partial x^J})^b n^{(I)} _b
=\delta ^I _J $ which is non-zero. Hence it is sufficient to require that $\delta A =0$ under
 deformations along the
 vector field $\vec{\xi} =
 \lambda \Sigma _I  h^{(I)} {\partial \over \partial x^I}$, $\lambda \ll 1$,
where the $h^{(I)}$ are arbitrary smooth functions. 

The deformation under $\xi$ defines a new surface $\cal{B}_\lambda$
parameterized as $\xI =\lambda \xi ^I (u^\beta ) , x^\alpha =u^\alpha +\lambda \xi^\alpha (u^\beta )$,
where the $u^\beta$ are coordinates on the surface.
To find the new area, one needs to compute
$\det g|_{\lambda } =\det g|_0 (1+ B^{ab} tr\delta g_{ab}|_\lambda  )$, where
evaluation on $\cal{B}_\lambda$ is indicated by the index $\lambda$. There are two
contributions to $\delta g_{ab}|_\lambda$, coming from the change in the metric components
and from the change in the differentials. Substituting into the general form of the metric
(\ref{nearmetricapp}), one finds
\begin{equation}\label{minimal}
ds^2 |_\lambda - ds^2 |_0 =2 \lambda \left( \xi^b \partial _b B_{\mu\alpha}
+2\partial _{(\mu}\xi ^\beta g_{\alpha ) \beta} \right) du^\alpha du^\mu
\end{equation}

Hence $\sqrt{-\det g}|_{\lambda } -\sqrt{-\det g}|_0  =\sqrt{-\det g}|_0 
({1\over 2}B^{\alpha\beta} \cal{L}_\xi B_{\alpha\beta} )$. Since this must be true for arbitrary
fields $\vec{\xi }$, we find the generalization of the codimension one condition,
that $B^{\alpha\beta} {\cal L}_\xi B_{\alpha\beta}  =0 $. In the case that 
$unit$ normal vectors  are well behaved, this becomes the condition that the
two independent extrinsic curvature tensors are traceless on the minimal surface,
$B^{\alpha\beta} \ki_{\alpha\beta}=0$, see equation (\ref{pitok}). For a block diagonal
metric, this means that $B_{\alpha\beta}$ must be quadratic in the
transverse conformal coordinates $x^I$ about the bubble at $x^I =0$.

\section{Open Questions}

A  calculation for future work is to require that the metric (\ref{blockmetric}) is a solution to
 the vacuum Einstein equation, with the correct boundary conditions on $\bub$ to
 give distributional curvature. It would be interesting to see if 
to see if there are other solutions than the known symmetrical cases. This might be
facillitated by using the equations in the form (\ref{curvaturetwo}). A second technical
issue is to drop the restriction of a block diagonal metric. In dynamical  situations,
there will certainly be cross-terms in the metric. Does this destroy the 
model of concentrated curvature? Is topology important, so that compact $\bub$
are stable against dynamics, whereas for an infinite cosmic string one
is forced to a finite width description?

An important issue is whether a codimension two submanifold
can have concentrated stress-energy that is different from a cosmological constant
on $\bub$.  Recall that this  form arises because with our assumptions
(listed before equation (\ref{intcurv}) ) only the scalar curvature contributes
to the components of the distributional Einstein tensor that are tangent to $\bub$.
To get a different equation of state, the Ricci tensor must also contribute 
  in (\ref{stressenergy}). Could this occur? A delta--function type contribution
  to the integral arises from a term in which two derivatives in the normal direction act on a function,
  and then choosing the function to be appropriately singular.
  This structure occurs in the scalar curvature of the transverse metric $R[\sigma ]$, 
  and yields the results of this paper. Inspection of (\ref{project}) shows that there are no
  other terms with this structure. If one wants  the metric on $\bub$  to be smooth, 
this means one has to look at other components of the metric $g_{I\alpha}$, which are not tangent
to $\bub$. Hence one must study metrics that are not block diagonal.
Turning to the more general Gauss-Codazzi relations in (\ref{gcrels}), one
finds that there do occur 
 terms with two normal derivatives on the functions  $g_{I\alpha}$. It would be
interesting to see if these terms can indeed give a distributional contribution to the
Ricci tensor on $\bub$.

Two types of questions that
 motivated this work  were to model branes wrapping cycles of compact submanifolds
and brane world world cosmologies. If the observed universe is confined to a brane
in a spacetime with ten  dimensions, then
one is interested in concentrating curvature on a
 submanifold with  codimension six.
Likewise, there are many solutions to ten and eleven dimensional
supegravity that represent $black$ branes wrapping cycles of submanifolds, but
one might ask if there are solutions that represent branes which are not collapsed
to black branes? These
configurations typically involve wrapping gauge potentials on submanifolds
with codimension greater than two.
For either of these problems, one needs a formalism which applies to curvature concentrated on 
surfaces with higher codimension than one or two. 

 It is not obvious that  there is a generalization to codimension greater than two.
The fact that the plane normal to $\bub$ is two-dimensional was used
heavily in the current construction.
Most significantly, the famous flat space minus a wedge metric was exploited, as
this metric has a curvature density that is a delta-function. Taking a solid angle out
of flat three dimensional space does not give a similar result--the curvature is not
focused. In three or more spatial dimensions the  expectation is  that
concentrated matter collapses to a black object. 
However, if our universe really is ten (or eleven) dimensional, it is 
worth pursuing these questions.

The relevance of these issues hinges on (i) how useful the thin-object model is 
in gravity--shells have been extremely useful, and (ii) 
how one thinks of branes in the context of classical gravity. Quantum mechanically,
particles are not points and a string is not a line. But what is the classical description
of a  stack of branes? In a classical gravitational calculation, must branes leap
from being test objects, existing on lower dimensional submanifolds,
 to being black branes? Or is there a middle ground, and if so, is the stack of branes thin
 in a classical sense?
 
  `I leave it to whomsoever it may
 concern, whether the tendency of this work be altogether to recommend quantum
 tyranny, or reward classical disobedience' \cite{Jane}.

\subsection*{Acknowledgements}I would like to thank David Kastor and Lorenzo Sorbo
for detailed discussions, and Narayan Menon and Benny Davidovitch for helpful
conversations about membranes. This work was supported in part by NSF grant PHY-0555394.

\appendix


\renewcommand{\thesection}{Appendix \Roman{section}}
 
      \section{Frobenius' Theorem and criteria to block diagonalize the metric}
      
 In general,  the metric has the form near $x^I =0 $
     \begin{equation}\label{nearmetricapp}
    ds^2 = g_{IJ}dx^I dx^J + 2g_{I\alpha}dx^I dx^\alpha 
    +B_{\alpha \beta}dx ^\alpha dx^\beta 
        \end{equation}
       Suppose that the mixed terms $g_{I\alpha}$ are nonzero. We want to find a new
        set of coordinates $ x ^{\prime a}$ such that
        \begin{equation}\label{diagone}
        0=g^{\prime\alpha I} ={\partial x^{\prime I} \over \partial x^c} 
        {\partial x^{\prime \alpha} \over \partial x^d} g^{cd} , \qquad I=1,2
        \end{equation}
        
        We look for a solution of the form $x^{\prime I} = x ^I$ and $x^{\prime \alpha}
        =F^\alpha (x^b )$. After some algebra, equations (\ref{diagone}) become
        \begin{equation}\label{diagtwo}
        \vecni  (F^\alpha ) \equiv g^{Ia} {\partial\over \partial x^a} (F^\alpha )=0 , \qquad I=1,2 ,
        \end{equation}
         where we have extended the definition of the normals  
         off the surface as vector fields using the spacetime metric,
            $n^{a(I)} =g^{Ia} {\partial\over \partial x^a}$.
         
         Frobenius' theorem addresses the question of whether there exist solutions $F^\alpha$.
          Assume that a specification of two smooth vector fields $\{ \vecni \}$ is given in a 
      neighborhood of the bubble.  Then Froebenius' Theorem states that
     the vector fields are integrable  if and only if the commutator
      of the vector fields closes,
     \begin{equation}\label{commute}
  [ \vec{n}^{(1)} ,  \vec{n}^{(2)}] = f^1  \vec{n}^{(1)} + f^2 \vec{n}^{(2)}
  \end{equation}
  for some functions $f^I$. 
        
       In order that solutions exit to (\ref{diagtwo}), it is necessary that the $\vecni$ satisfy
       the Frobenius Condition (FC), equation (\ref{commute}). For necessity,
       suppose that there is a solution $F^\alpha$ to (\ref{diagtwo}). Then it must also be true
       that $0=   [ \vec{n}^{(1)} ,  \vec{n}^{(2)}] F^\alpha $. However,
        if the vectors $\vecei$ do not satisfy the FC, then this statement is inconsistent. 
        
        For sufficiency, suppose that the $\vec{n}^{(I)}$ satisfy the FC.  Then Frobenius' theorem
        states that the $\vec{n}^{(I)}$ are a basis for the tangent space of 
        a smooth submanifold $N$. Choose good coordinates $y^I$ on $N$.
        Then the coordinate basis vectors ${\partial \over \partial y^I}$ form
         another basis for $T(N)$, and so each $\vec{n}^{(I)}$ is a linear combination of
         the ${\partial \over \partial y^I}$.  The equations (\ref{diagtwo}) become
         equivalent to the set ${\partial \over \partial y^I}F^\alpha =0$.  Choose
        $D-2$ other coordinates $y^\alpha$ , so that 
        $(y^I , y^\alpha )$ are good coordinates on $M$, in a neighborhood of $N$.
         Then the solutions to (\ref{diagtwo})
        are just functions which are independent of the coordinates on $N$, for example,
        $F^\alpha =y^\alpha$.
       
       Hence a necessary and sufficient condition that the metric can be  put
in block diagonal form is that the normal vectors satisfy the FC (\ref{commute}).

 \section{Gauss-Codazzi Details}   
 This appendix assembles various pieces of the Gauss-Codazzi formalism that are
 used in the text. Here it is assumed that the metric is smooth. We follow the notation of 
 \cite{RayPhDthesis}.
 
 The metric is split as in equation (\ref{metricnear}),
$g_{ab} =B_{ab} +\sigma_{ab}$. There are two independent extrinsic curvature tensors,
    
\begin{equation}\label{bigextrcurv}
\pi _{ab}^{\ \  c} =B_a ^m B_b ^n \sigma _d ^c \nabla _m B_n ^d \quad and \quad
\lambda _{ab}^{\ \   c} =\sigma_a ^m \sigma_b ^n B _d ^c \nabla _m B_n ^d
\end{equation}
$\pi _{ab}^{\ \  c}$ is orthogonal to $B$ in its third index, and  tangent to $B$ in its first two. The opposite
is true for $\lambda _{ab}^{\ \  c}$--it is tangent to $B$ in its third index, and orthogonal to $B$ in its first two.
The assumption that the surfaces $\bubeps$ are submanifolds implies that $\pi$ is symmetric
in its first two indices. $\lambda$ is symmetric only if $\sigma$ is the metric for a submanifold
as well (as is assumed in the body of the paper).b

The three index tensor $\pi$ can be made to look more like the familiar extrinsic curvature
for a codimension one submanifold as follows.
Let $\hatei$ be a set of orthonormal  normal forms on $\bubeps$, so that  $\sigma_{ab}$ can be expanded as $\sigma_{ab} =\Sigma _I \hatei _a \hatei_b$. Substituting into
the definition of $\pi$, and using $\nabla _m B_n ^d =- \nabla _m \sigma_n ^d$, one finds
\begin{equation}\label{pitok}
\pi _{ab}^{\  \  c} = -\Sigma_I K^{(I)}_{ab} \hat{e} ^{(I)c }, \quad where \quad
 K^{(I)}_{ab} =B_a ^c \nabla _c \hatei _b 
 \end{equation}
 
 We will need the Gauss-Codazzi relations for the projections of the Riemann tensor  \cite{RayPhDthesis}:
\begin{eqnarray}\label{gcrels}
B_a ^e B_b ^f B_c ^m B_d ^ n R_{efmn} [g]  &=R_{abcd} [B] - \pi_{acr} \pi_{bd} ^{\ \  r} 
+ \pi_{bcr} \pi_{ad} ^{\ \  r}  \\   \nonumber
\sigma_a ^e B_b ^f \sigma_c ^m B_d ^ n R_{efmn} [g]  &=
\sigma_a ^e B_b ^f \sigma_c ^r B_d ^ n \left( \nabla _e \pi _{fnr}  -\nabla _f \lambda _{ern} \right)
-\lambda _{eab} \lambda ^e _{\  cd}  -\pi _{eba} \pi ^e _{\  dc} \\  \nonumber
\sigma_a ^e \sigma_b ^f \sigma_c ^m \sigma_d ^ n R_{efmn} [g]  &=R_{abcd} [\sigma ]
 - \lambda_{acr} \lambda_{bd} ^{\ \  r} 
+\lambda_{bcr} \lambda_{ad} ^{\ \   r}  
\end{eqnarray}

 These imply the following relation for the scalar curvatures,
 \begin{equation}\label{curvatureapp}
 R[g] =R[B] +R[\sigma ] + \pi_a \pi^a -\pi_{abc} \pi^{abc}  + \lambda_a \lambda^a
  -\lambda_{abc} \lambda^{abc} +2(\nabla_a \pi^a -\nabla_a \lambda^a )
 \end{equation}
 where $ \lambda_a =\sigma^{mn}\lambda_{mna} ,  \pi_a =B^{mn}\pi_{mna}$. $R[s]$
 is the scalar curvature for the metric $s_{ab}$, that is, the curvature of the 
 derivative operator that annhilates $s_{ab}$
 Substituting(\ref{pitok}) gives equation (\ref{curvature}).

 \end{document}